%% file: main.tex
\documentclass{IEEEcsmag}

\usepackage[colorlinks,urlcolor=blue,linkcolor=blue,citecolor=blue]{hyperref}

\usepackage{upmath}

\newcommand{\lmttfont}{\fontfamily{lmtt}\selectfont}

\usepackage{graphics}
\usepackage{color, colortbl}
\usepackage{graphicx}
\usepackage[table,xcdraw]{xcolor}
\usepackage{booktabs} 

\usepackage{setspace}

\newcommand{\rev}[1]{{\textcolor{black} {#1}}}

\usepackage{amsmath}%
\usepackage{MnSymbol}%
\usepackage{wasysym}%

\newcolumntype{L}[1]{>{\raggedright\arraybackslash}m{#1}}
\newcolumntype{C}[1]{>{\centering\arraybackslash}m{#1}}
\newcolumntype{R}[1]{>{\raggedleft\arraybackslash}m{#1}}

\definecolor{LightCyan}{rgb}{0.88,1,1}
\definecolor{lightgray}{gray}{0.9}

\usepackage{ragged2e}

\usepackage{verbatim}

\begin{document}


\title{Virtualization over Multiprocessor System-on-Chip: an Enabling Paradigm for Industrial IoT}

\author{Alessandro Cilardo, Marcello Cinque, Luigi De Simone, Nicola Mazzocca} 
\affil{Dept. of Electrical Engineering and Information Technology at the University of Naples Federico II, Via Claudio 21, Napoli, Italy}









\begin{abstract}
\justifying
The next-generation Industrial Internet of Things (IIoT) inherently requires smart devices featuring rich connectivity, local intelligence, and autonomous behavior. Emerging Multiprocessor System-on-Chip (MPSoC) platforms along with comprehensive support for virtualization will represent two key building blocks for smart devices in future IIoT \rev{edge} infrastructures. We review \rev{representative existing solutions}, highlighting the aspects that are most relevant for integration in IIoT solutions. \rev{From the analysis, we derive} a reference architecture for a general virtualization-ready edge IIoT node. We then analyze the implications and benefits for a concrete use case scenario and identify the crucial research challenges to be faced to bridge the gap towards full support for virtualization-ready IIoT nodes.
\end{abstract}

\maketitle

\begin{figure*}[!t]
\centering
\includegraphics[width=1.95\columnwidth]{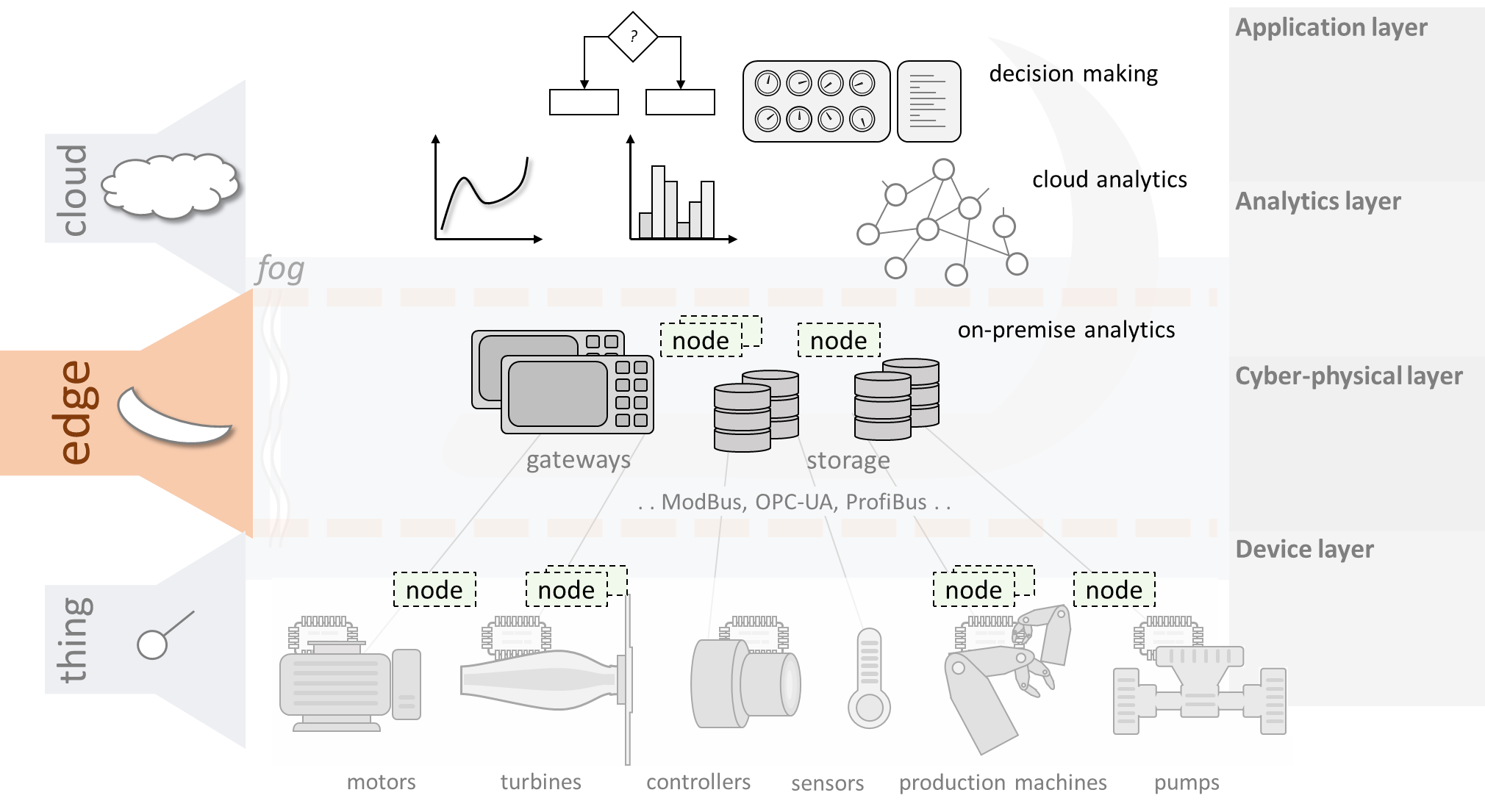}
\caption{The IIoT landscape. \rev{High-end devices at the \emph{edge} level is the focus of this article.}}
\label{fig_IIoT}
\end{figure*}

\chapterinitial{Industrial Internet of Things} represents a key pillar for the emerging revolution of Industry 4.0 in a large spectrum of areas, from smart factories to energy infrastructures and process automation, which require smart devices featuring rich connectivity, local intelligence, and autonomous behavior. The landscape of IIoT technologies is depicted in Figure~\ref{fig_IIoT}.
\rev{These can be mapped to the \emph{thing}, \emph{edge}, or \emph{cloud} layers, depending on their scope and physical closeness to plant systems. In particular, the devices, or \emph{nodes} placed at the edge level play the fundamental role of a glue technology between the things (e.g., smart sensors and/or actuators) and the cloud. Heavyweight \emph{on-premise} analytics, distributed supervisory control, massive real-time data collection/storage, and, in general, any task not bearable directly by thing devices can be \textit{offloaded} to the edge.} 
Nodes at this level involve a continuum of technologies, rather than a discrete separation of levels, as confirmed by the emerging paradigm of Fog computing~\cite{ieee18-ref}. 
\rev{The envisioned trend} is to move control, communication, computation, and decision-making across the boundaries of the infrastructure between the central cloud and the network edge, closer to the location, i.e. \rev{the things}, where data is generated.

In this scenario, \rev{edge} nodes must meet stringent requirements, such as performance scalability, interoperability, agility and reconfigurability for efficient deployment, dependability attributes (e.g., safety), timeliness, and \textit{mixed-criticality}, i.e., running applications with different levels of criticality on the same hardware, beyond traditional regulatory requirements involving functional safety (IEC61508) and control (IEC61131).

\rev{These desiderata are simply out of reach for traditional programmable logic controllers currently used in industrial environments. For this reason, we are witnessing the emerging adoption of MPSoC in industrial scenarios~\cite{youness14}.}
MPSoCs feature asymmetric multiprocessing (AMP) with dedicated hardware accelerators and rich I/O capabilities, which are crucial to let edge nodes withstand IIoT requirements.

\rev{As a specific example of a mixed-criticality workload at the edge level, let us consider a wind turbine plant, needing real-time control as well as monitoring of significant amounts of electrical and mechanical parameters for predictive maintenance with machine learning (ML), which has crucial economic implications because of the very high Operation and Maintenance (O\&M) costs of such systems~\cite{mohammadi2020, acromag}.
We envision that different functionalities, ranging from real-time control to ML algorithms, with their diverse needs and requirements, can \textit{cohabit} in the same mixed-criticality edge node platform. For instance, high-speed data acquisition can be performed using dedicated hardware accelerators, which are tightly coupled with the controller tasks requiring highly deterministic and safe behavior.
These, in turn, may share the same computation resources of predictive maintenance tasks, running online ML algorithms to classify the current state of the machinery (e.g. mining signatures hidden in physical parameters, providing hints for anomalies or possible faults) and predict possible maintenance interventions.}
Being all in one, the smart device can be used to control different machines, simplifying deployment and maintenance tasks in highly automated future smart factories.

\rev{To fully enable such a scenario on top of MPSoC hardware platforms, we advocate the crucial need of \textit{virtualization} to create isolated, flexible, and safe execution environments on next-generation IIoT edge nodes.  
Virtualization is well established in cloud infrastructures} to abstract hardware resources by creating \textit{isolated application environments}, called Virtual Machines (VMs) or guests, running \rev{operating systems} and applications on the same physical machine. The \textit{hypervisor} (or Virtual Machine Monitor) is the software layer that implements this abstraction and manages the resources on behalf of VMs, exposing virtual resources to them, such as virtual CPUs (vCPUs) and virtual memory, that are mapped to physical resources.

In this paper, we outline the main opportunities provided by emerging MPSoC technologies for the IIoT area, along with \rev{an analysis} of current hypervisor solutions \rev{and how these technologies can converge in a reference architecture for virtualization-ready IIoT edge nodes.}
\rev{From the analysis}, we find that these hypervisors \textit{need to be revisited} to fit the features of MPSoCs and fully match IIoT requirements, as they only offer limited support to properties of relevance~\cite{ruh2019need}, including:

\begin{itemize}

    \item \textit{Determinism}: the hypervisor must provide consistent mechanisms to guarantee predictable latencies for critical VMs;

    \item \textit{Safety Certification}: for safety-critical applications, the hypervisor must strictly meet industry regulations and standards;

    \item \textit{Isolation}: the hypervisor must fully guarantee isolation of VMs in terms of timeliness, security, and safety, \rev{despite the use of underlying shared hardware resources};
    
    \item \textit{Flexibility}: the hypervisor must provide effective means to replace VMs at runtime for full reconfigurability and scalability.

\end{itemize}

Matching the above challenges with virtualization-ready MPSoC platforms will have disruptive implications for IIoT in next-generation Industry 4.0 \rev{edge} infrastructures. 

\section{Virtualization for IIoT Edge Platforms}
\label{sec-edge}

\rev{
Today's embedded and IoT technologies offer unprecedented opportunities for edge IIoT applications.
}
Systems on Chip (SoC), particularly MPSoC platforms, introduce a brand-new design dimension, beyond the PLC and low-end devices dominating the traditional realm of industrial applications.
\rev{Such systems normally feature multi-processor parallel architectures supporting complex processing and/or redundancy, dedicated acceleration functions, either with fixed functionality, like GPUs, or fully customizable at the hardware level, as well as rich equipment of standard communication interfaces for cost-effective, direct integration in IIoT infrastructures.}
%
%
%
%
%
As a flavor of current MPSoC technologies, field-programmable gate arrays (FPGAs) have evolved from mere glue logic solutions to large-scale reconfigurable hardware platforms, used both in embedded and data-intensive applications requiring high-performance levels.
%
%
New industrial applications for FPGAs include mechatronics, robotics, and power systems, mostly involving digital real-time simulation, advanced control techniques, and electronic instrumentation~\cite{rodriguez15-ref}. Like GPUs, among other tasks, FPGAs are particularly suitable for accelerating machine learning operations. 

Emerging MPSoC platforms, featuring AMP capabilities and FPGA/GPU acceleration can play a key role in addressing the expected mix of requirements in terms of connectivity, scalable computing power, interoperability, manageability, as well as safety and security.
%
At the same time, such platforms also pose increased difficulties at the programming level.
Heterogeneous cores, including general-purpose processors, real-time processors, GPUs, and FPGA-implemented soft-cores, require remote processor management and messaging, while suitable application programming interfaces and abstractions are needed for the device access, interrupt handling, memory management, synchronization, etc.
The support for AMP in IoT-class systems has only recently been addressed by the technical community.
Further, unlike general-purpose server-class architectures, full virtualization support in IoT platforms has been explored only recently, particularly by ARMv8 architecture.
For embedded general-purpose processors, virtualization requires a large range of features such as hardware support for vCPUs and instruction emulation, consistent mapping of hypervisors to privilege levels,
e.g. allowing type-2 hypervisors, two-level memory address translation and address space identifiers, virtual interrupt management, and advanced features such as nested virtualization.
For heterogeneous MPSoC platforms featuring GPU and FPGA acceleration, virtualization poses even tougher challenges.

\input{virtualization_solutions.tex}

\begin{figure*}[!t]
\centering
\includegraphics[width=2\columnwidth]{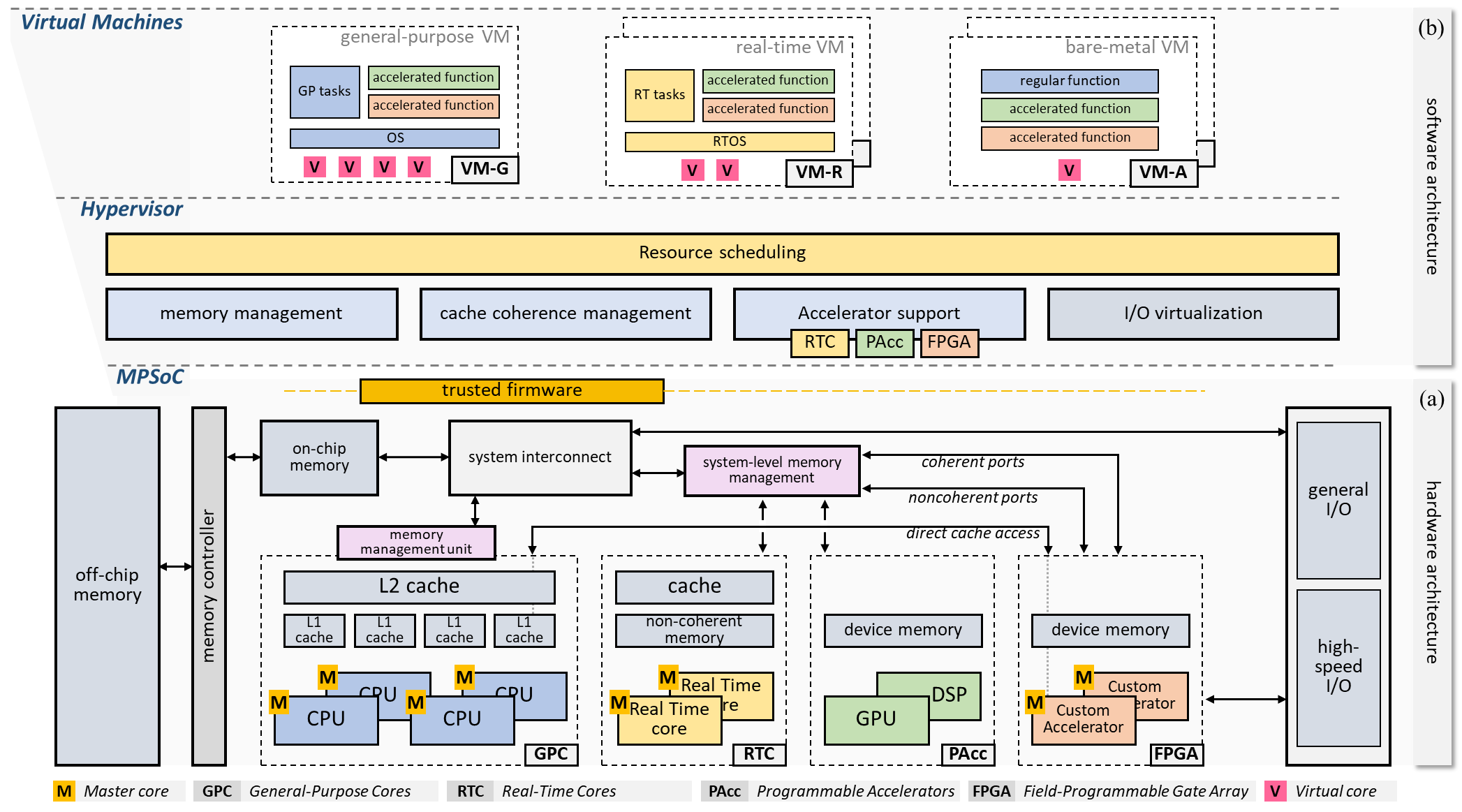}
\caption{IIoT \rev{edge} node reference architecture}
\label{fig_refArch}
\end{figure*}

\section{A Reference Architecture for Virtualization-ready IIoT \rev{Edge Nodes}}
\label{sec-ref}


Figure~\ref{fig_refArch} depicts the detail of a reference architecture for an IIoT node platform.
We first focus on the hardware layer, then move to the software stack,
in line with the technological opportunities summarized in the previous section.

\subsection{Hardware Architecture}
\rev{
Part (a) in Figure~\ref{fig_refArch} focuses on the key building blocks of the hardware portion in the architecture.
In addition to general-purpose 32- or 64-bit cores, featuring full support for virtual memory, caching, complete operating systems, etc., these blocks include
}
\begin{itemize}

    \item \textbf{Real-Time Cores (RTCs)}, providing non-coherent predictable memory, lightweight memory-protection units, low-latency interrupts, lock-step execution mode, and other functions aimed at safety and performance predictability.

    \item Loosely coupled \textbf{programmable accelerators (PAcc)} such as GPU and DSP units, offering bulk processing power to certain classes of workloads, typically for non-critical application components.

    \item \textbf{Dedicated accelerators}, possibly implemented on an FPGA hardware-reconfigurable fabric, which may either act as passive accelerators or as customized \emph{master} cores accessing system resources such as high-performance ports to the system interconnect as well as caches for coherent communication between heterogeneous cores.

    \item A parallel, high-performance \textbf{system interconnect} along with a complex memory management infrastructure, mediating system-wide memory accesses from a variety of heterogeneous subsystems.

    \item Special \textbf{support for security} features and \textbf{execution of trusted firmware} at the highest privilege level, for secure boot-time configuration and establishment of a \emph{chain of trust} across the system software stack.

    \item A rich set of \textbf{integrated Input/Output controllers}, such as SPI, CAN, UART, USB, etc., connected both to the general system interconnect and to the hardware-reconfigurable fabric, which can feature direct low-latency access to external data streams.

    \item \textbf{Integrated controllers to the external memory}, e.g. high-capacity DDR banks.

    \item \textbf{A monitoring infrastructure} for controlling physical parameters such as temperature, voltage, power, etc, both for reliability and for security purposes, e.g. to detect tampering.
\end{itemize}

\subsection{Software architecture}


Emerging hypervisors targeted at MPSoC platforms are faced with nontrivial technical challenges. \rev{According to the previously identified characteristics of hypervisors suitable for IoT/IIoT edge nodes, we claim that a solution} should include at least the following key features, depicted in Figure~\ref{fig_refArch}.(b):

\begin{itemize}
    
    \item \textbf{Resource scheduling}: the hypervisor has to assign master cores to VMs, in the form of \textit{virtual cores}, and schedule them considering real-time constraints, such as time budget and deadlines. Either fixed time partitioning or resource reservation approaches can be adopted. Access to the memory and I/O resources must be regulated as well, as discussed next.
    
    \item \textbf{Memory management}: the physical memory is shared among VMs and virtual cores. The hypervisor must ensure both address space isolation, through the system-level memory management, and time isolation, ensuring a predictable maximum memory stall time for each VM, e.g., by statically assigning memory access slots to VMs.
    
     \item \textbf{Accelerator support}: RTCs, GPUs, and FPGA-based accelerators are shared hardware resources and their assignment to VMs must be carefully addressed by the hypervisor, also considering isolation constraints. Accelerators could also be seen as statically assigned virtual cores, such as an RTC core or a softcore on the FPGA dedicated to a VM. 
    
    \item \textbf{Cache Coherence management}: current MPSoC architectures foresee multiple cache levels shared among cores and accelerators; in addition, if the MPSoC platform allows direct access to caches from accelerators, applications running on a CPU core may experience a severe impact in terms of the timing behavior, as even the accelerators, in addition to other cores, can evict cache blocks. Cache coloring should be supported and extended by the hypervisor to ensure isolation.
    
    \item \textbf{I/O virtualization}: the hypervisor must allow consistent and isolated access to I/O controllers. Many solutions are available, such as full emulation, paravirtualized I/O, pass-through, and hardware-assisted virtualization, which often require full support to IOMMU.
    
\end{itemize}

The all-in-one features of smart IIoT devices can be conceptually implemented with three types of VMs with different requirements running within the same MPSoC platform on top of the hypervisor, enabling a mixed-criticality environment. In Figure~\ref{fig_refArch}.(b), we highlight:

\begin{itemize}
    \item \textit{VM-G}: a non-critical VM running a GPOS. This VM may for example be dedicated to predictive maintenance tasks, running on a full-fledged OS on general-purpose cores with all required machine learning libraries. The VM can be interconnected to cloud services, sharing data and plant models constructed from past observations on similar devices.
    \item \textit{VM-R}: a critical VM running an RTOS. This VM can be dedicated to the real-time control of the plant, using an RTOS and virtual cores mapped on RTCs.
    \item \textit{VM-A}: a special-purpose VM running \emph{bare-metal} software needed to interface accelerated functions which, in turn, perform online data acquisition and signal processing at high speed (e.g. through dedicated hardware functions programmed on FPGAs).
\end{itemize}
The hypervisor assures that the different VMs can accomplish their objectives without interference. Different combinations are possible according to the specific application needs.

\subsection{Use Case Scenario}
\input{use_case}

\section{Challenges and Open Issues}
\label{sec-chal}

\rev{
Along with the mapping to the reference architecture, the above case study highlights inherent needs and requirements of emerging IIoT, ranging from time predictability to security, which have to be effectively addressed by future virtualization-ready platforms targeted at industrial environments.
Driven by this observation, we identified a set of technical challenges that must be answered to fully achieve the scenarios we described in this paper.}

\noindent
$\blacksquare$ \textbf{Determinism}.
The hypervisor must be designed to assure predictive behavior to VMs that host hard real-time applications. Classical solutions, e.g., schedulability analysis, partitioning, and worst-case execution time (WCET) analysis approaches need to be extended to address \rev{the new sources of non-determinism introduced by AMP in current MPSoCs}, such as taking into account accelerators and their impact on memory, caches, and WCET.

\noindent
$\blacksquare$ \textbf{Certification \& Testing}. Safety-critical systems raise challenges from the certification point of view since almost all standards recommend V\&V activities to provide a safety integrity level (SIL). Studies in the literature \cite{cotroneo2018dependability} and international standards provide guidelines for testing activities, which encompass fault injection, robustness, and performance testing, error impact analysis, coding standards, code review, etc. 
These tasks are performed at the different levels of the virtualized architecture, considering the use of Commercial-Off-The-Shelf (COTS) components, requiring a review of original test suites and a redesign of the approaches traditionally used in the industry. Some hypervisors provide support for certification, limited in some cases to a given domain (e.g., avionics for Xtratum). Redesign of the certification package is still needed if targeting new platforms or different domains. A general solution for certification is not yet available for the IIoT.




\noindent
$\blacksquare$ \textbf{Mixed-criticality}. The reference architecture allows in principle the realization of mixed-criticality IIoT nodes, embedding functions with different criticality levels on the same hardware. The hypervisor shall implement criticality-aware solutions, e.g., provide the highest isolation degrees to highly critical VMs. Fault and attack isolation needs to be enforced and proven according to standards. 

\noindent 
$\blacksquare$ \textbf{Multi-mode virtualization}. Real-time and functional safety devices combine multiple platforms into a single product via virtualization, to meet the demand from the embedded market of mixed-criticality functionalities. Emerging embedded virtualization solutions try to address the requirements for IIoT platforms, supporting hardware-assisted virtualization, partitioning, and paravirtualization I/O for increasing the performance of critical devices. Unfortunately, no virtualization solution can address all these multi-mode virtualization aspects.

\noindent
$\blacksquare$ \textbf{Support for heterogeneous acceleration}. Virtualization support for accelerators is fairly immature. In server environments, there is some support, especially for GPUs, consisting either in API remoting, fixed/mediated pass-through (limited in the degree of flexibility they ensure), or device emulation approaches (overly expensive in terms of performance overheads). None of these approaches is directly supported at the hardware level by embedded GPU devices. Some studies classify virtualization support for server-class FPGAs in terms of resource-level, node-level, and multi-node level approaches~\cite{Vaishnav2018}. However, FPGA devices feature a different coupling between the CPUs and the hardware-reconfigurable fabric and cannot just re-use node/multi-node approaches. Resource-level virtualization is not a mature technique for FPGAs and the potential performance overheads it may cause make its role questionable in both embedded and high-end settings.

\noindent
$\blacksquare$ \textbf{Orchestration}. \rev{New solutions are emerging for \textit{lightweight} orchestration, such as k3s (\href{https://k3s.io}{https://k3s.io}), specifically targeting IoT and edge devices. However, the support to determinism (e.g., time-sensitive placement and migration of resources), to AMP (e.g., taking into account shared accelerators in deploy decisions) and to mixed-criticality (e.g., performing criticality-aware VM deployment) is still overlooked by current solutions.}

\noindent
$\blacksquare$ \textbf{Security}. Digital security is a major concern for tomorrow's IIoT applications. The main challenge lies in the complexity of their hardware/software architecture, making it difficult to establish trustworthiness in the whole supply chain, from the silicon up to the software stack. Indeed, cybersecurity threats may impact the device and the whole industrial infrastructure. Future solutions need to consider standards, like ISO/IEC 15408, which impose different security requirements, e.g., partitioning and resource isolation levels, full control over communication channels, and the development of auditing mechanisms. 

\section{Conclusion}
\label{sec-conc}
We discussed a reference hardware/software architecture for next-generation IIoT \rev{edge} nodes. Current MPSoC platforms and embedded hypervisors already provide promising features in terms of resource management, accelerator support, and determinism. However, considerable research and development efforts are still needed from the research community to achieve consistent virtualization technologies, both at hardware and software levels, fully addressing the open challenges towards future virtualization-ready IIoT nodes.

\bibliographystyle{IEEEtran}
\bibliography{IEEEabrv,bibliography}



\end{document}

%% file: virtualization_solutions.tex
\subsection{MPSoC Virtualization Solutions}
\label{sec:io_virt_solutions}

\input{solutions_table}


\rev{Driven by the challenges posed by MPSoC platforms at the software level, as highlighted above, we briefly review the most representative approaches to MPSoC virtualization, particularly those potentially suitable for IoT/IIoT.}
While an exhaustive analysis is covered by past published surveys \cite{ramakrishnan2020comprehensive,lee2018impact},
we highlight the most significant features that researchers and practitioners have identified for virtualization in the domain, to shape an ideal IoT/IIoT hypervisor in our reference architecture, fully matching the opportunities brought by MPSoC platforms for AMP applications.
\rev{In particular, we analyzed solutions offering MPSoC platform support (based on ARM architecture in most cases), small codebase (for certification purposes), real-time capabilities (temporal isolation), memory protection (enhanced cache coherence), I/O isolation (e.g., IOMMU support), I/O virtualization flexibility (e.g., paravirtualized devices, fully-emulated devices, etc.), acceleration support (FPGA, GPU, etc.).}


\rev{\textbf{Xen} \cite{XEN_Barham2003} is a well-known hypervisor that \textit{paravirtualizes} VMs, i.e., the guest needs to be modified to communicate directly with it, through \textit{hypercalls} (i.e., the PV mode). Actually, Xen can use also hardware emulation (i.e., the HVM mode) and both full-virtualization and paravirtualization features (i.e., the PVHVM mode).} 
Xen features a Real-Time Deferrable Server (RTDS) scheduler to offer guaranteed CPU capacity to guest VMs, which is fundamental for time isolation. Xen also provides a \textit{null scheduler}, that allows static assignment of vCPUs to physical CPUs (pCPUs), removing any scheduling decision. Xen has been used as a building block for Xilinx systems \cite{xen_ultrascale}, including the \textit{System Memory Management Unit} (SMMU) and full control of memory partitioning for AMP applications. \textit{Cache coloring} \cite{cache_coloring} improves determinism since VMs get their allocation of cache entries with no shared resources. 
%
%
FPGAs have also been integrated as devices in Xen \cite{wang2013pvfpga}, showing that accelerators can be shared among multiple guest OSes.


\rev{Other approaches exploit Trusted Execution Environment (TEE)-based virtualization, mainly provided by ARM. \textbf{LTZvisor} and \textbf{RTZVisor} \cite{pinto2019demystifying} are representative ARM TrustZone-based solutions that implement \textit{dual-guest OS} and \textit{multi-guest OS} scenarios, respectively. The general-purpose OS (GPOS) will run within the \textit{non-secure world} and a real-time OS (RTOS) in the \textit{secure world}, assuring time and memory protection. Although the LTZVisor hypervisor family fully supports some ARM FPGA-based boards, no information is provided about the seamless access to accelerators. ARM TrustZone provides a dual MMU and cache interface, thus cache partitioning is guaranteed only for dual-guest OS configurations, requiring cache coloring (not implemented) for multi-guest OS configurations.}


\textbf{Jailhouse} \cite{jailhouse} is a Linux-based partitioning hypervisor, enabling AMP applications to cooperate with the Linux kernel to run bare-metal applications \rev{and multiple guest OSes}. The main purpose is to enhance isolation rather than provide classic virtualization; thus, the hypervisor splits CPUs, memory, devices, etc., into \textit{cells}, i.e., strongly isolated domains, each of them assigned to one guest OS, and its applications, called \textit{inmates}. 
Recently, in the context of the Hercules H2020 project, page coloring is supported in Jailhouse. Jailhouse supports accessing FPGA accelerators from cells thanks to OpenAMP's Remoteproc/RPMsg and VirtIO technologies.
%
%

\rev{\textbf{Xtratum} \cite{masmano2009xtratum} is a paravirtualized partitioning hypervisor, certifiable for the avionic domain since it is ARINC 653 compliant. Xtratum supports several CPU families (including ARM), providing strong temporal isolation by exploiting a fixed cyclic scheduler. Partitions' memory is allocated statically, with no memory areas shared between them. Xtratum defines a minimum set of deterministic hypercalls and enables interrupts for the running partitions, to minimize temporal interference among partitions. Cache coherence is enforced with \textit{non-cacheable} memory areas for the virtual memory map. Despite FPGA virtualization is not currently supported, Xtratum is being evolved to serve as a building block in next-generation space industry devices that use this kind of accelerators \cite{hermes_project}.}



\textbf{ACRN} is an open-source lightweight bare-metal hypervisor for IIoT scenarios and edge device use cases, which leverages hardware-assisted CPU extensions for spatial and temporal isolation \cite{li2019acrn}. It considers user or service VMs and real-time VMs (RTVM), i.e., a special isolated user VM optimized to offer real-time capability with dedicated physical resources. Memory partitioning is assured with hardware-assisted virtualization (e.g., Intel EPT, VT-d). ACRN provides rich embedded I/O virtualization, including GPU virtualization mediated by pass-through techniques provided by Intel GVT-d. However, there is no explicit support for FPGA virtualization.

\textbf{Bao} \cite{martins2020bao} is a lightweight bare-metal hypervisor designed for mixed-criticality IoT systems. Bao focuses on security and safety requirements by providing strong isolation, fault-containment, and real-time features. When a VM is instantiated, Bao statically partitions the resources with no scheduling intervention, since it maps each vCPU to a single pCPU. The memory is statically assigned exploiting two-stage translation hardware virtualization support, while I/O virtualization is pass-through only. Bao implements page coloring to mitigate interference between guest partitions due to contention at shared last-level caches (LLCs) \cite{cache_coloring}. 

Table \ref{related_works_table} summarizes the main features of the analyzed hypervisors. We considered three classes for hypervisor size according to Lines of Code (LOC), namely, \emph{Small} (less than $10$kLOC), \emph{Medium} (less than $100k$LOC) and \emph{Large} (greater than $100k$LOC) classes. As expected, all solutions are mostly small (in terms of LOC) and pay special attention to vCPU scheduling, often adopting static solutions to achieve strong isolation. Memory and I/O management (via Input/Output MMU - IOMMU), allowing static assignment of resources, is quite mature in most solutions. However, not all hypervisors implement mechanisms for handling cache coherence, which represents a source of non-determinism. Furthermore, explicit support for accelerators is missing in most cases, highlighting that current solutions are only partially ready for IoT/IIoT use cases.
\color{black}
We used the features reported in the table to evaluate the {\lmttfont IoT/IIoT-readiness} indicator, as they are related to the IoT/IIoT requirements described in the previous section, as follows:

\begin{itemize}
    
    \item \textit{Determinism}. This crucial requirement for real-time systems is linked to {\lmttfont Scheduling};
    
    \item \textit{Safety certification}. No solutions are certified (some are certifiable) according to safety-related standards, and the number of LOC (i.e., {\lmttfont Size}) directly impacts the certification effort;
    
    \item \textit{Isolation}. Includes both time and space isolation, linked to {\lmttfont Scheduling}, {\lmttfont Memory Management}, {\lmttfont Cache Coherence}, and {\lmttfont I/O Isolation};
    
    \item \textit{Flexibility}. It is directly related to {\lmttfont I/O Virtualization} and {\lmttfont Accelerators Support}, as they simplify the management of I/O devices for embedded systems (e.g., through VirtIO, full emulation, or pass-through) and the transparent sharing of FPGAs, GPUs, or other accelerators.
    
\end{itemize}
Using the above mapping, we assigned the {\lmttfont IoT/IIoT-readiness} (level 1 to 5) in Table \ref{related_works_table} to indicate the level of coverage of the hypervisor's features over the requirements mentioned before: level 5 indicates that the solution is fully ready for IoT/IIoT applications, while level 1 suggests that the hypervisor lacks real applicability. For example, despite providing strong isolation, the Xtratum hypervisor presents a very low IoT/IIoT-readiness level since it currently does not provide any support for IOMMU and accelerators (this is work in progress \cite{hermes_project}); rather, both ACRN and Jailhouse hypervisors can be considered nearly ready for IoT/IIoT scenarios due to consistent support for accelerators (both FPGAs and GPUs) and I/O virtualization, which are critical points for MPSoC platforms.
\color{black}

%% file: solutions_table.tex
\begin{table*}[hbt!]
\caption{\rev{Relevant features of the analyzed virtualization solutions for IoT/IIoT edge nodes.}}
\label{related_works_table}
\resizebox{\textwidth}{!}{%
\sffamily 
\footnotesize 
\setstretch{0.90}
\rowcolors{1}{}{lightgray}
\begin{tabular}{C{1.5cm}C{1.5cm}C{3cm}C{3cm}C{2cm}C{2.3cm}C{2.5cm}C{2.5cm}||C{1.5cm}}
\toprule\toprule
\rowcolor{LightCyan}
\textbf{Hypervisor} & 
\textbf{Size} & 
\textbf{Scheduling} & 
\textbf{Memory Management} & 
\textbf{Cache Coherence} & 
\textbf{I/O Isolation} & 
\textbf{I/O Virtualization} & 
\textbf{Accelerators Support} & 
\textbf{IoT/IIoT-readiness}
\tabularnewline
\midrule
    Xen &
        Large (x86, x86-64), Medium (ARM) & 
        Real-Time Deferrable Server (RTDS), \textit{null} scheduler (partitioning) & 
        Hardware-assisted, Direct paging, Shadow Paging &
        Cache coloring &       
        ARM64 SMMU, Intel IOMMU &                   
        Emulated, Paravirtual drivers, Pass-through (Intel VT-d), SR-IOV & 
        Server-class FPGA support \cite{wang2013pvfpga} & 
        \includegraphics[width=0.08\columnwidth]{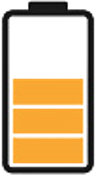}
        \\ 
    
    LTZVisor, RTZVisor & 
        Small & 
        Static allocation of CPU cores (dual-guest), Cyclic scheduling (multi-guest) & 
        Hardware-assisted (ARM TZASC and TZPC) &  
        ARM TrustZone dual MMU and cache interface (multi-guest scenario is not supported) & 
        Support ARM64 SMMU & 
        Pass-through policy &
        N/A & 
        \includegraphics[width=0.08\columnwidth]{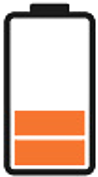}
        \\ 
        
    Jailhouse &  
        Small & 
        Partitioning & 
        Static Partitioning & 
        Cache coloring (work in progress) & 
        ARM64 SMMU, Intel IOMMU & 
        Pass-through (Intel VT-d) &
        OpenAMP and VirtIO support & 
        \includegraphics[width=0.08\columnwidth]{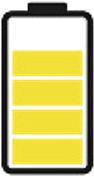}
        \\ 
    
    Xtratum &            
        Small & 
        Fixed cyclic scheduler &  
        Static Partitioning &              
        Set memory areas uncacheable in the config file &       
        N/A &                   
        Pass-through, VirtIO  &
        Under development (HERMES2020 project \cite{hermes_project} &                    
        \includegraphics[width=0.08\columnwidth]{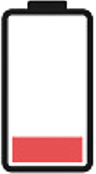}
        \\ 
    
    ACRN & 
        Medium & 
        CPU core partitioning algorithm &
        Hardware-assisted (Intel EPT) &
        N/A &
        Intel IOMMU & 
        Full emulation, Pass-through (Intel VT-d), VirtIO  &
        Support to virtual GPU through Intel GVT-d &
        \includegraphics[width=0.08\columnwidth]{lev_4.png}
        \\ 
    
    Bao &  
        Small &
        Static Partitioning &
        Static Partitioning &
        Cache coloring &
        ARM SMMU work in progress &
        Pass-through  &
        N/A &
        \includegraphics[width=0.08\columnwidth]{lev_3.png}
        \\ 
       \bottomrule

\end{tabular}
}
\end{table*}

%% file: use_case.tex
\begin{figure*}[!t]
    \centering
    \includegraphics[width=2\columnwidth]{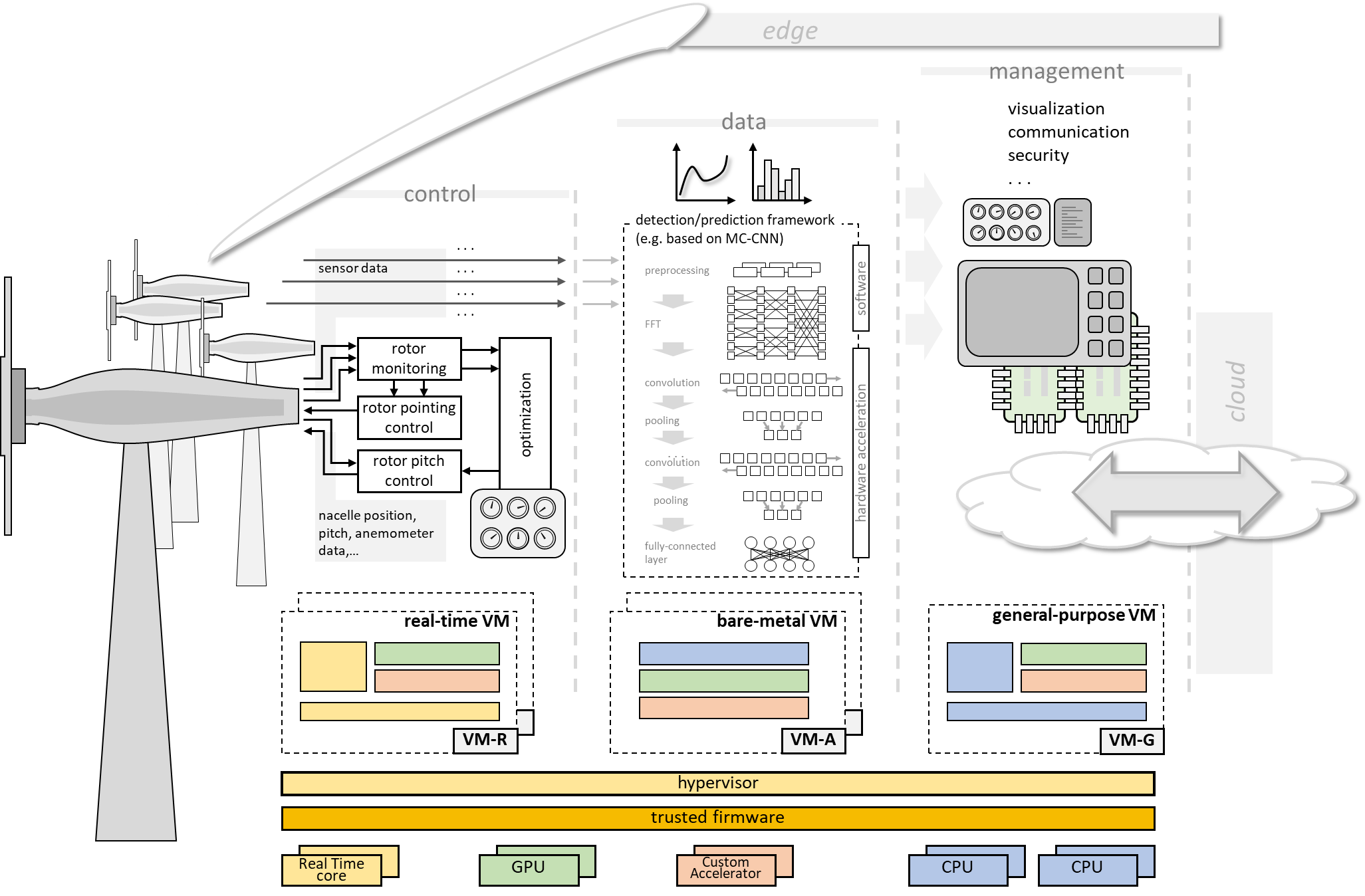}
    \caption{A wind turbine plant use case for the reference architecture discussed in this article}
    \label{fig_useCase}
\end{figure*}

To illustrate the above reference architecture, we refer to a representative case study focusing on control and monitoring of wind turbine plants.
These rely on Supervisory Control and Data Acquisition (SCADA) systems to support condition monitoring, tracing physical parameters such as ambient, nacelle and stator winding temperature, voltage, current and generation efficiency, harmonics in line, etc.
While a variety of mechanical and electrical sensors supply massive streams of data, hints to detect anomalies or faults are revealed by signatures hidden in electrical or other physical signals. Emerging machine learning approaches are suitable to identify such hints and are a promising pathway for predictive maintenance.
However, while high-level decision-making and maintenance strategy is performed in a centralized way, it is desirable that pre-processing and refinement of collected data for analytical purposes are performed locally, as close as possible to the data sources, while only high-level information is conveyed to the centralized infrastructure.
\rev{An MPSoC-based hardware-reconfigurable device is an ideal fit as an edge node in such an infrastructure can control and monitor wind turbines and support on-premise data analytics.
Figure~\ref{fig_useCase} depicts the mapping between the reference architecture described in the previous section and the components of the wind turbine case study,
highlighting the impact of the virtualization-ready MPSoC edge node and the role of the different hosted VMs.
The architecture enables programming flexibility, allowing fast reconfiguration of the monitoring system, yet deterministic timing capabilities, relying on high-performance controllers and direct I/O processing through dedicated FPGA acceleration.
Specifically, multiple real-time VMs deployed in the MPSoC handle control loops involving the direct operation of the turbine~\cite{acromag}, e.g. enforcing a certain target pitch and pointing of the turbine to achieve a target RPM for optimal energy generation, while exploiting reserved virtualized resources for meeting strict timing constraints and accessing dedicated physical interfaces, e.g. dozens or hundreds of single-ended I/O points typically featured by medium-end MPSoC platforms.
A dedicated acceleration VM, on the other hand, is devoted to speed up on-field live analysis of complex high-resolution data,
autonomously identifying hazardous conditions and/or supplying refined data to the remote control system.
For example, this scenario naturally fits the potential of FPGA-based MPSoC platforms,
where compute-intensive tasks are accelerated through hardware-implemented blocks
(e.g. the different layers of multi-channel convolutional neural networks, used as a deep learning framework for condition monitoring and fault detection~\cite{mohammadi2020}, as suggested in Figure~\ref{fig_useCase}),
while less demanding tasks like sensor data preprocessing are performed in software by the virtualized CPUs made available to the acceleration VM.
}
Last, a non-critical general-purpose VM can simultaneously manage other functions, e.g. alarm monitoring and communications over networks.
In that respect, the hardware-based security features available within the MPSoC, such as tamper resistance and support for trust chains in the software stack, also play a crucial role in a variety of possible cyber-attacks, an important concern for emerging IIoT architectures.

